\documentstyle[12pt,aaspp4]{article} 

\def\gsim{\ \raise -2.truept\hbox{\rlap{\hbox{$\sim$}}\raise5.truept
\hbox{$>$}\ }}
\def\msun{M_{\odot}} \def\mincir{\ \raise -
2.truept\hbox{\rlap{\hbox{$\sim$}}\raise5.truept	\hbox{$<$}\ }}
\def\magcir{\ \raise -2.truept\hbox{\rlap{\hbox{$\sim$}}\raise5.truept
\hbox{$>$}\ }}
\begin{document}

\title{The SZ Effect as a Probe of Non-Gravitational Entropy in
Groups and
Clusters of Galaxies} \author{A. Cavaliere$^1$ and N. Menci$^{2}$}
\affil{$^1$
Astrofisica, Dip. Fisica 2a Universit\`a, Roma I-00133} \affil{$^2$
Osservatorio
Astronomico di Roma, Monteporzio, I-00044}

\def\lsim{\ \raise -2.truept\hbox{\rlap{\hbox{$\sim$}}\raise5.truept
\hbox{$<$}\ }}
\def\gsim{\ \raise -2.truept\hbox{\rlap{\hbox{$\sim$}}\raise5.truept
\hbox{$>$}\ }}
\def\msun{M_{\odot}}

\begin{abstract}

We investigate how strongly
and at what scales the Sunyaev-Zel'dovich effect reflects the shifting balance between
two processes that compete for governing the
density and the thermodynamic state of the hot
intra-cluster medium pervading
clusters and groups of galaxies. One such process is
the hierarchical clustering of the DM; this induces gravitational heating
of the diffuse baryons, and strives to push 
not only the galaxy systems but also 
the ICM they contain toward
self-similarity. 
Away from it drives the other process, constituted by
non-gravitational energy fed back into the ICM by the condensing
baryons.
We base on a semi-analytic model of galaxy formation and clustering
to describe how the baryons are partitioned among the hot,
the cool and the stellar phase; the partition shifts as the galaxies
cluster hierarchically, and as feedback processes (here we focus on stellar winds and
Supernova explosions) follow the star formation. 
Such processes provide a moderate feedback, 
whose impact is amplified by the same 
large scale accretion shocks
that thermalize the gravitational energy of gas falling into the 
growing potential wells. 
We use the model to compute the Compton parameter $y$ that governs the 
amplitude of the   
SZ effect, and to predict how this is affected by the feedback;  
for individual clusters and groups we find a relation of $y$ with the ICM temperature, the $y-T$ relation, 
which departs from the form suggested by the self-similar scaling,  
and bends down
at temperatures typical of
galaxy groups. We also compute the average $\langle y\rangle$
and the source counts as a function of $y$ under different assumptions
concerning feedback strength and cosmology/cosmogony.
We then discuss to what extent our results are generic of 
the hierarchical models of galaxy formation and clustering; 
we show how the $y - T$ relation, to be measured at $\mu$wave or sub-mm
wavelengths, is model-independently related to the shape of the $L-T$ correlation
measured in X-rays.
 We conclude that these observables
together -- because of their complementarity and their observational
independence -- can firmly bound the processes responsible for
non-gravitational entropy injections into the ICM.
\end{abstract}


\section{Introduction}

The evolution of galaxies and galaxy systems may be understoood in terms 
of a competition between two processes. One is the dominant gravitational drive of the
dark matter (DM) that leads to hierarchical clustering;
this builds up ``haloes'' that grow
larger and deeper by the merging of smaller clumps (see Peebles 1993), and yet
remain closely {\it self-similar}, i.e., with central densities and
profiles which are nearly scaled
versions of each other (see Navarro et al. 1995). The other process
is the active
{\it response} of the baryons contained in such potential wells; this 
strives to
drive the radiative structures {\it away} from self-similarity.

Groups and clusters of galaxies provide an ideal testing ground for such a
picture. On the one hand, their deep gravitational wells are dominated by DM
masses ranging from $M\approx 10^{13}$ to $10^{15}~M_{\odot}$. Their sizes 
may be 
defined by the virial radius $R$ where the DM density $\rho$ at formation
exceeds by a factor of about $2\,10^{2}$ the background value 
$\rho_u(z)\propto (1+z)^3$; they 
scale as
$R \propto (M/\rho_u)^{1/3}$, and match or exceed $R\sim 1.5 \, h^{-1}$ Mpc
for
the rich clusters. The well depths may be measured in terms of 
the circular  velocities
$v = (G\, M/R)^{1/2} \propto
M^{1/3}
\rho_u^{1/6}$, which attain 
or exceed 1400 km s$^{-1}$ in rich clusters today.

On the other hand, such wells also contain large masses of diffuse baryons,
of
order $M_b \approx 0.15\, M$ (White et al. 1993; White \& Fabian 1995;
Fukugita,
Hogan \& Peebles 1998); these are in the state of a hot plasma, the
intra-cluster medium or ICM.

Its temperatures may be estimated on assuming also the ICM to be in virial
equilibrium; this provides the scaling $kT \approx kT_{V}\propto v^2$, and values in
the
range $kT \sim 0.5  \div 10$ keV in going from poor groups to rich clusters. By
and
large, these results are confirmed by the observations of many
high-excitation
lines at keV energies and of the X-ray continuum.

In fact, the ICM is well observable at X-ray energies through its thermal
bremsstrahlung emission $L\propto  n^2\,T^{1/2}\, R^3$, which from groups to rich
clusters ranges from $ 10^{42}$ to several $10^{45}$ erg s$^{-1}$ for typical
densities
$n\sim 10^{-(3\pm 1)}$ cm$^{-3}$.

But the same  hot electrons also inverse-Compton upscatter the photons of
the
CMB, and thus they tilt slightly the black-body spectrum toward higher 
energies; at microwave wavelengths 
the result toward individual clusters is an apparent intensity decrement $\Delta I/I \approx - 2\,y$ 
in terms of the Compton parameter $y\propto  n\,kT\,R$.
Numerical values
$|\Delta I/I| \sim 10^{-4}$ had been predicted  by Sunyaev \& Zel'dovich 
(1972)
(hence the acronym SZ effect), and are being measured in a growing number of
rich clusters; for reviews and recent data, see Rephaeli
(1995), Birkinshaw (1999), Carlstrom et al. (2000), Saunders \& Jones (2000).
The computation of $y$ and of related observables from a comprehensive 
model for
the thermal state of the ICM constitutes the main scope of the present paper; 
its plan is as follows. In \S  2 we discuss how  the SZ probe 
fits into the current debate concerning the 
role of non-gravitational processes in setting the
thermal state of the ICM.
In \S 3
we describe how we compute such state, and how our model fits into
the
current efforts toward including star formation and other 
baryonic processes in the hierarchical
clustering picture; at the end of the section we show how the SZ observables are
computed on the basis of our specific model. In \S 4 we present our results 
and predictions concerning the
$y-T$
relation and the integrated SZ observables, namely, the source counts and
the
average Compton parameter $\langle y \rangle$. In \S  5 we discuss the
building
blocks of our model, and make contact with other related works; 
we also stress the generic aspects of the link between the $y-T$ and the 
$L-T$ relations. 
In the final \S
6 we summarize our predictions and conclusions. 

\section{The State of the ICM}

Before we embark into technical issues, we discuss the framework of
observations and theories in which the present work is to fit.

\subsection{The self-similar picture}

The simplest picture of the ICM would have its thermal energy  to be 
of purely gravitational origin; that is to say, 
originated only by gravitational energy released on
large scales comparable with $R$ as the external gas falls into the
self-similar potential wells provided by the DM (Kaiser 1986), and
thermalized by
some effective coupling. By itself, the process would establish temperatures
$T\approx T_V$ and produce self-similar ICM distributions in the wells,
with
central densities $n$ just proportional to the DM density $\rho$, hence to the
external
$\rho_u$. If so, the DM scaling recast into the form $R\propto
T_V^{1/2}
\, \rho^{-1/2}$ easily yields that the resulting X-ray luminosity
should scale as 
\begin{equation} 
L_{grav} \propto \rho_u^{1/2}\, T_V^2.
\end{equation}

However, this simple picture is challenged --  especially at the scales 
from poor clusters to groups of galaxies -- by several recent X-ray 
observations. 

\subsection{Broken self-similarity}

First, the observed shape of the $L-T$ correlation departs from
$L\propto  T^2$ and bends down progressively in moving from very 
rich clusters down to poor groups (see
Ponman et al. 1996; Helsdon \& Ponman 2000), as if lower $T$  should imply
considerably lower values of $n$ and thus much lower luminosities $L$.
Second,
the profiles of the X-ray surface brightness do not appear to be
self-similar,
rather they flatten out at low $T$ (Ponman, Cannon \& Navarro 1999). Third,
the
X-ray luminosity function $N(L, z)$ shows little or no evidence of
cosmological
evolution (Rosati et al. 1998); if anything, weak negative evolution may
affect
the bright end of the LFs, but surely the data rule out the strong, positive
``density evolution'' that would prevail if the relation 
$n \propto \rho$ held and kept  $L(z)$ roughly constant. 

All that points toward an ICM  history more complex than pure,
gravitational heating from infall into the forming DM potential wells, as
already
realized by Kaiser (1991). It rather suggests additional 
heating by injection of some 
non-gravitational energy, which ought to be more
important for the shallower wells of the groups to the point of  
breaking there the simple assumption $n \propto \rho$.

To single out the baryonic processes responsible for this, it is clearly
useful (see Cavaliere 1980, and
Bower 1997) to consider the ICM specific entropy (in Boltzmann units) 
$S =  ln\,
p/n^{5/3} \propto ln\, kT/n^{2/3}$, the one combination of $T$ and $n$ that
is
invariant under pure adiabatic compressions of the gas falling into the
wells.
Data and N-body simulations (Lloyd-Davies, Ponman \& Cannon 1999) concur to
show
that in moving from clusters to groups this quantity is far from behaving in
the
way suggested by the self-similar scaling 
\begin{equation} 
S_{grav}\propto
ln\,T_V \, \rho_u^{- 2/3}~. 
\end{equation} In fact, when the values averaged
out to $10^{-1}\, R$ are plotted as a function of $T$, they  level off to a
``floor'' corresponding to $e^S \propto k T/ n^{2/3} \approx 100$ keV
cm$^2$. On
the other hand, in clusters $ S$ as a function of $r$ increases
outwards
(see Lloyd-Davies, Ponman \& Cannon 1999). These behaviours indicate that
the entropy
is enhanced in two ways relative to the self-similar scaling, both in the
present groups and during the history of the clusters.

\subsection{Non-gravitational processes}

One promising explanation of all these observational features goes back to
heating/ejection of the ICM by thermal/mechanical feedback effects from 
star formation during
the process of galaxy build up and clustering into groups 
(Cavaliere, Menci \& Tozzi 1997, 1999;
Wu, Fabian \& Nulsen 1998; Valageas \& Silk 2000; Fujita \& Takahara 2000).

We shall adopt as a baseline the injections of momentum and of thermal 
energy
contributed both by stellar winds (Bressan,
Chiosi \& Fagotto 1994) and by Type II Supernovae (SN) that follow the 
formation of massive stars in 
shallow wells. The latter include not only the current groups, but also the
progenitors
of current clusters; in fact, the  hierarchical clustering paradigm
envisages
such events to take place also during the early phase of a cluster's history
when the progenitors were just groups, which subsequently grew by
the
progressive inclusion of additional mass lumps of comparable or smaller
sizes.

Such combined feadback effects naturally imply the ICM to be not only
pre-heated
to about $0.2 \div  0.3$  keV, but also to be dynamically ejected outwards to the
perifery of the DM haloes, at low densities $n\sim 10^{-4}$ cm$^{-3}$. 
This is how a relatively high entropy level corresponding to $kT/n^{2/3} \sim 10^2$ keV
cm$^2$
may be attained in groups as well as in the central regions of the forming
clusters. In the latter, the subsequent evolution will include adiabatic
compression of the gas accreted along with the DM into the deepening wells
(Tozzi \& Norman 2000), a process that conserves entropy. But the gas
falling
supersonically  into such deep potential wells will also pass through strong
accretion shocks at about the virial radius $R$, the process whereby the ICM
is
actually heated to the current high temperatures of order $T_V$. By the same
token, much additional entropy is deposited in the outer regions,
corresponding
to shock conversion of the bulk inflows $v_1 \approx v$ into thermal
energy.

This scenario leads one to expect in the groups not only lower densities,
and
much lower bremsstrahlung emission compared with the scaling $L \propto
T^2$, as 
observed; but it also leads to expect little cosmological evolution for the
population of bremsstrahlung emitters constituted by groups and clusters.
This is because in the  
X-ray luminosity function $N(L, z)$ the ``density evolution''
driven at medium-low $L$ by the hierarchical clustering is
balanced by the lower luminosities corresponding to a steeper $L-T$
correlation.

In fact, we have investigated and computed elsewhere in detail (Menci \&
Cavaliere 2000, hereafter MC2000; Cavaliere, Giacconi \& Menci 2000,
hereafter
CGM2000) how the stellar feedbacks (heating combined with
ejection) affect the ICM and its X-ray emission $L\propto n^2\, R^3 \, T ^{1/2}$,
which is particularly sensitive to changes of the central ICM density.

\subsection {The SZ probe}

In the present paper we stress the {\it complementary} probe provided by the
SZ effect. The CMB modulation
$\Delta I/I \approx -2\, y \propto n\,kT\,R $
introduced in \S 1 provides another observable combination of $T$ and $n$ 
which depends
on the thermal and on the mechanical feedback in a more balanced way.

In parallel with the scaling for $L$ provided by eq. (1), 
the self-similar scaling for $y$ reads 
\begin{equation} y_{grav} \propto
\rho_u^{1/2}\, T_V^{3/2}~.  
\end{equation} 
We shall investigate how the actual 
$y$ also {\it departs} away from $y_{grav}$ due to $n\neq \rho$ 
and $T\neq T_V$ holding at groups scales, and shall discuss how
the relative relative strength $y/y_{grav}$ of the SZ signal is  related
to entropy. Our investigation will start with computations based 
upon specific semi-analytic modelling for the
baryons in groups and clusters, in particular for those diffuse and hot
constituting the ICM.

\section{Semi-analytic Modelling of the ICM State  and Evolution}

In the scenario outlined in \S2.3 the processes responsible for the
non-gravitational entropy floor in groups in 
clusters are
just the same as those governing the formation history of stars and
galaxies. At
galactic scales, they tend to suppress the star formation in the earlier and
shallower potential wells. The observable consequences include 
flattening of 
the local, optical LF function at faint-intermediate luminosities, and a 
decline of the 
integrated star formation rate for $z\gtrsim 2$; but the degree of
such a decline sensitively depends on the feedback
strength in the relevant wells.

The picture has been quantitatively implemented in the so-called
semi-analytic
models  for star and galaxy formation (SAMs, see Kauffmann et al. 1993; Cole
et al. 1994; Somerville \& Primack 1999). These are built  upon 
the notion of galaxies as baryonic cores located 
inside DM haloes (White \& Rees 1978); they  specify -- though often in a
phenomenological form -- how the baryons are cycled among the condensed
stellar
phase, the cool gas phase and the hot gas phase, under the drive of the
merging events that make up  the hierarchical evolution of the DM haloes.

Since the very same DM and baryonic processes also play a key role 
in determining the state of the ICM on scales larger than galactic, we take up the SAM approach
in its latest versions, and use it to evaluate the
amount of non-gravitational heating of the ICM and to predict
the corresponding SZ effect. Our model is necessarly complex as it 
comprises several processes and many computational steps, so we outline 
its frame and workings in the form of a flow chart in fig. 1.

\subsection{Our SAM: the DM and stellar sectors}

We base on an updated version of the SAM originally presented 
in Poli et al.
(1998).  In this approach  the average over the many possible merging
histories
of a given structure is performed 
by convolution over the merging probability functions 
given by Lacey \& Cole
(1993)
and tested, e.g., by Cohn, Bagla \& White (2000); 
the average also includes the 
convolution
with the probabilities for galaxy aggregations driven by dynamical friction inside
common haloes. 
Toward computing the stellar observables, the model 
treats the following processes.

1) The DM haloes merge following the hierarchical clustering. The detailed
merging
histories are important for the problem at hands, since the gas carrying the
non-gravitational entropy is 
pre-heated/expelled from condensations smaller
than present-day systems, and is subsequently accreted by larger haloes. The
time delays between heating/expulsion and accretion propagate up the 
structure 
hierarchy the aftermaths of the feedback.

2) Part of the hot gas  mass $m_h$ diffused throughout the haloes 
cools down in all differently-sized potential wells, then 
condenses into stars on the timescale $\tau_*$, see Table 1.   
The amount of cool gas is also controlled by the stellar feedback. 
This is because
the condensation in stars of a mass $\Delta m_*$ eventually causes 
stellar winds and SN explosions to release a total non-gravitational energy 
$ E_* = E_{SN} \, \eta_{SN}\, \Delta m_*$ ergs, where 
$E_{SN}\sim 10^{51}$ erg is the energy 
of a Type II SN explosion, $\eta_{SN}\approx 3.2\, 10^{-3}/M_{\odot}$ 
is the efficiency in SNe for the Scalo IMF. 
Part of such energy is coupled to the gas 
and reheats an amount of cool gas $\Delta m_h$;   
in the SAMs this is often related 
to $\Delta m_*$ by the phenomenological assumption $\Delta m_h
=\Delta m_*\,(v/v_h)^{-\alpha_h}$, with typical values for  
the fraction of reheated gas  
given by $f_*\equiv \Delta m_h/m_h\sim 10^{-1}$.  

The importance of the thermal effects of the feedback 
 is expressed in terms of the above parametrization by 
 the ``stellar'' temperature 
\begin{equation}
k T_* = (1-q_0)\,E_{*}\,m_p/3\, \Delta m_h~, 
\end{equation} 
where $q_0\lesssim 10^{-1}$ is the fractional energy 
going directly into bulk motion of the material;  
$KT_*$ is to be compared with the virial temperature $k\, T_V$.  
In the way of a preliminary estimate, note that 
$k T_* = 0.7 \,\Delta m_*/\Delta m_h \gtrsim 0.2$ keV holds 
for a stellar baryonic fraction exceeding $1/5$, the low value 
actually appropriate only 
for rich clusters.
But $kT_V \approx 0.2$ keV is also the virial temperature corresponding 
to $M\sim 5\, 10^{12}\, M_{\odot}$; so in  galaxies  
and in poor groups 
the stellar heating is expected to cause the gas to flow out of the wells.  
In the numerical model the process is evaluated in detail for all haloes.  

In fact, the scale dependence of the feedback is marked 
by the key parameter $\alpha_h$. To identify its    
relevant range consider the ratio between the 
work done in moving the gas out of the well and 
the actual bulk kinetic energy,  given by 
$\epsilon_0 = f_*\, m_h \,
v^2/q_0\,E_* \propto
(v/v_h)^{2-\alpha_h}$.   
The proportionality constant turns out to be around $0.5$ for our fiducial 
values $E_{SN}=5\,10^{50}$ erg, 
$\eta_{SN}=5\,10^{-3}/M_{\odot}$ (with the contribution from stellar winds 
included after Bressan et al. 1994) and for $q_0=0.1$; 
so the reheated gas does escape from the relevant haloes with 
circular velocities $v\geq v_h$ when one takes $\alpha_h\geq 2$, but 
it constitutes a fraction $f_*$ that decreases with increasing $v$. 

In the model development presented 
by MC2000 and CGM2000 we allowed this parameter to range from the extreme value
$\alpha_h=5$ adopted by other workers to our fiducial value $\alpha_h=
2$. The former value (hereafter case $\cal{A}$) yields $\epsilon_0
\propto v^{-3}$, which in the shallow early wells constitutes quite a 
{\it strong} 
feedback, well over and above the escape energy barrier; 
this leads
to considerable depletion of the cool baryons 
and causes the star formation rate to decline sharply
for $z\magcir  2$.
The latter value (hereafter referred to as case $\cal{B}$) instead yields a
{\it moderate} and ``neutral'' feedback corresponding to $\epsilon_0 =$ 
const;  
thus even the shallow, galactic haloes
are allowed to retain considerable amounts of gas, and the resulting star
formation rate is high and rather flat for $z \magcir  2$.

The values of this and of other relevant parameters are collected 
in Table 1. These lead to the B-band luminosity functions of the galaxies 
and to the cosmic star formation rate presented  
in MC2000, when the critical universe dominated by Standard 
Cold Dark Matter is adopted. In the 
canonical $\Lambda$-dominated flat universe the same set of parameters leads 
to the 
results presented by Poli et al. (1999) together with 
the related galaxy sizes and Tully-Fisher relation, and discussed by CGM2000.

\subsection{Our SAM: the ICM sector}

The SAM used in the present paper 
includes also the processes relevant to the ICM that have
been developed and implemented in MC2000 and CGM2000. In those papers the baryonic component
condensed into stars has been used to normalize the model parameters to the
optical observations, and the cool gas component has been 
presented for a final check with relevant data; but our
main focus was onto the {\it hot} gas identified with the ICM, 
and onto its
X-ray emission $L$. 
The hot ICM gives also rise to the SZ effect of interest here, 
so the enumeration begun in \S 3.1 is continued next, to cover also 
the following processes relevant to the present scope. 

3) The accretion of external gas into the potential wells of groups and clusters
sets the boundary condition for the internal gas disposition.
To this aim, we
compute the density and temperature of groups and clusters at their 
virial radius $R$. This is done in terms of the external 
temperature $T_1$ of the gas to be accreted during the merging histories 
of groups and clusters; $T_1$  is 
set not only by the virial equilibrium inside the merging lumps 
but also by the feedback, and the more strongly so the closer 
to $T_*$ are the virial values.  
Our  computation comprises the following steps. 

First, we assume that the DM density $\rho(r)$ 
inside the cluster has the 
form given by Navarro, Frenk \& White (1997), and that the density 
$n_1$ external to the virial radius $R$ is related to the baryon 
density $\Omega_b$ by $n_1=\rho(R)\, \Omega_b$. 

Next, for a group or cluster of mass $M$ we compute 
the temperature $T_1$ of the external infalling gas. 
In particular, for each lump of mass $M'$ accreted onto $M$
we consider two contributions: 
the fraction $1-f_*(M')$ remains inside the potential well of 
$M'$ at the virial temperature $T_V(M')$, while 
the fraction $f_*(M')$ is ejected 
by the winds and the SNe gone off within $M'$, and is heated to 
the temperature $T_*$ given by eq. (4). 

Last, we assume that the external gas is incorporated into 
the mass $M$ passing through a shock front at a position 
close to the virial radius $R$. 
This is expected because the 
latter separates the region which is in hydrostatic equilibrium 
from the outer regions where the motion is dominated by infall 
(see also Bower et al. 2000),  
and so defines the transition from infall kinetic to 
thermal energy of the gas. 
Shock positions close to $R$ are confirmed 
by many 1-D and 3-D  N-body simulations (e.g., 
Knight \& Ponman 1997; Takizawa \& Mineshige 1998; Pantano, Gheller \& Moscardini
1998; Governato et al. 2001, in preparation). 

Then we compute the density $n_2$ and the temperature $T_2$ of the ICM 
at the boundary $R$ of a group or cluster from 
the jump conditions at the shock; these 
are derived from mass and momentum 
conservation across the shock (the classic Rankine-Hugoniot conditions), 
and are expressed here in terms of the 
cluster potential $\phi_2$ at the boundary, and of the temperature  
of the incoming gas. The thermal energy $k\,T_1$ associated with 
an accreted lump $M'$ 
is contributed by the virialized component at $T_V'$ and by the 
ejected component at $T_*'$, with their weigths $1-f_*'$ 
and $f_*'$ introduced above.
Thus, the weighted density jump $G\equiv n_2/n_1$ at the shock in the ICM 
of the considered group/cluster of mass $M$ is given by 
\begin{equation}
G(M,M')=f_*' \,g(T_*')+(1-f_*' )g(T_V');
\end{equation}
the density jump for either  component has the 
form (Cavaliere, Menci \& Tozzi 1999)
$g(t)=2\,(1-t)+[4\,(1-t)^2+t]^{1/2}$ in terms of variable 
$t=T_*'/T_2$ or $t= T_V'/T_2$, respectively. 
The full expression for $T_2$ is given by 
the same authors (see their eq. 8), but for strong shocks it reads simply
\begin{equation}
k\,T_2(M)\approx {{\mu m_p \phi_2(M)}/3} + 3k\,T_1/2~, 
\end{equation}
in the approximation of small bulk kinetic energies downstream.  
Note that for given $M'$ the weighted jump $G(M)$ increases with $M$, 
but  saturates to the value 4 when  
$M\gg M'$, that is, for strong shocks. 

For a cluster or group of mass $M$ we obtain the {\it average} 
values of the weighted density jump 
and of the boundary temperature $T_2$ by integrating over the merging 
histories; that is, we sum 
over all accreted lumps $M'$ with their relative merging probabilities 
by inserting the above expressions eqs. (4) -- (6) in the semi-analytic 
context. 
This will introduce dispersion in the boundary conditions due to the 
intrinsic spread of the merging histories; we recall 
(Cavaliere, Menci \& Tozzi 1997, 1999) 
that the scatter we predicted  
for the $L - T$ correlation agrees with 
the findings by Markevitch (1998) and by Allen \& Fabian (1998). 

We have numerically checked that for weaker and weaker feedback 
the average of $G$ over the merging histories goes to a constant, so  
that the self-similar scaling $L\propto T^2$ is recovered as expected.

4) The final process relevant to the ICM is the 
hydrostatic equilibrium, from which we compute the internal gas density 
profile, given the 
above boundary conditions. The gas density 
run $n(r)/n_2$ 
is related to the DM density profile $\rho(r)$ in terms of the parameter 
$\beta = \mu m_H \sigma_2^2/kT_2\approx T_V/T_2$, 
the ratio of the DM to the gas specific energy 
($\sigma_2$ is the 1-D 
velocity dispersion of the DM at the boundary, related to 
$\phi_2$). While in the case of isothermal ICM and of constant $\sigma (r)$ the 
run is simply given by 
$n(r)/n_2 = \rho^{\beta}(r)/\rho_2^{\beta}$ 
(Cavaliere \& Fusco Femiano 1976, Jones \& Forman 1984), here we use the 
expressions given by Cavaliere, Menci \& Tozzi (1999) for 
the potential of  Navarro et al. (1997) and for a polytropic, but nearly 
isothermal ICM. 
\footnote{Because 
$T(x)$ is observed to be close to constant, we distinguish 
$T_V,\, T_2$ and
$T$
(the latter may be taken to be the
emission-weighted temperature) only when relevant.}
Thus, for a mass $M$ at the redshift $z$
we compute the inner ICM profiles $n(x)/n_2$ and $T(x)/T_2$
in terms of the normalized radius $x=r/R$. This is achieved basically 
with no free parameters
(see \S 5 for a discussion) 
since the profiles are  defined 
in terms of the boundary conditions and of the   
depth of the DM potential wells. 
In particular, for rich clusters $T_2\gg T_1$ holds and  
the shocks are strong, so the expression of $\beta$ obtains simply 
on using the expression 
(6) for $T_2$ to 
obtain  $\beta\approx 1/(1+3/2\,T_1/T_V)$.  
On the other hand, for decreasing $M$ the value of $\beta$ decreases 
to attain values $\approx 1/2$ for small groups, which 
yield gas density profiles significantly 
shallower than in clusters in agreement with Ponman, Cannon \& Navarro
(1999).

Fig. 1 provides a representation and a summary 
of the processes and of the computational steps we use to set the boundary 
conditions and the internal structure of the ICM. 
Our assumptions and fiducial choices are discussed in \S 5. 

We refer to our previous papers MC2000 and CGM2000 for added details and for
the results concerning the X-ray observables. In short, these include:
$\beta$ decreases and the relation $L-T$ bends down in
moving toward systems of lower $T$, while  $N(L,z)$ shows little or no
cosmological evolution, all in good agreement with the
observations. In addition, the counts of X-Ray sources from the ICM are
predicted to be enhanced corresponding to high values 
of the star formation rate at $z \magcir 1.5$.

Here we epitomize the effectiveness of our model by showing in fig. 2 the
central ICM entropy as a function of the X-ray temperature $T$.  
Such entropy is given by the ratio 
$T(r)/n(r)^{2/3}$ and is computed on 
using the profiles discussed above. Ultimately, these 
are determined by the global
gravitational and by the feedback-induced energies associated 
with the gas in the infalling clumps. 

The result is compared with
the data presented by Ponman, Cannon \& Navarro (1999) and by Lloyd-Davies, Ponman
\& Cannon (2000). We recall that entropy is of special value as a state
variable for the ICM; in fact, in conditions of long cooling times it keeps the
archive - as it were - of the thermal events (different from adiabatic compressions)
which are associated with the ICM gravitational heating/compression and with the
non-gravitational heating/ejection.

\subsection{The SZ effect from our SAM}

Here we focus on how we use the above model to compute the SZ effect from the
ICM in groups and clusters. The basic relations will be just
recalled, having been
covered in many excellent reviews beginning with Sunyaev \& Zel'dovich
(1980) and including Rephaeli (1995) and Birkinshaw (1999).

The Compton parameter induced by the ICM column
displaced from the center by the distance $w$ on the plane of the sky  may be
written as
\begin{equation} 
y(w, M) = -
2\,{\sigma_T\, k \over m_e\,c^2}\,\int d l\,T(r)\,n(r)= -2\,{\sigma_T\over
m_e\,c^2}\,kT_2\,n_2\,R\,\int dx\,{x\over \sqrt{x^2-w^2}}\,
{T(x)\over T_2}\,
{n(x)\over n_2} ~,
\end{equation}
where $\sigma_T$ is the Thompson cross-section, and $dl$ is the path element
through the structure along the line of sight. Our SAM 
provides $T_2$, $n_2$ at the boundary and
the
profiles $T(x)/T_2$, $n(x)/n_2$ all depending on the structure mass $M$,
as described in \S 3.2. We
extend the integrations to infinity, but take into account the temperature
and
density jumps at $r=R$ across the shock.

The SZ distortion of the CMB spectrum at a frequency $\nu$ is given by
\begin{equation}
\Delta I/ I = y\,j(\xi)~.
\end{equation}
The function $j(\xi)$ depending on $\xi \equiv h\nu/kT_B$ is
given by Sunyaev \& Zel'dovich
(1980) in terms of the temperature $T_{B}$ of the CMB black-body spectrum;
$j \approx -2$ holds in the Rayleigh-Jeans limit $h\nu \ll kT_{B}$,
appropriate for $\mu$wave observations. On the other hand, $j (\xi)$ becomes
positive beyond $\nu \approx 220$ GHz; so the SZ effect offers a second chance to measure
or check $y$ in the sub-mm band, which is clean of
radiosources
and is now coming of age with the recent or planned sub-mm instrumentation.

The signal from a individual structure at a
redshift $z$ in a detector beam with effective aperture $\theta_b$ and
angular response $\psi(\theta/\theta_b)$
produces  the equivalent flux
\begin{equation}
S_{\nu}=2 \, {(kT_{B})^3\over
(hc)^2}\, J(\xi)\,2\,\pi\,\int_0^{\infty}
d\theta\,\theta\, \psi(\theta/\theta_b)\, y(\theta) =
J(\xi) \; \Bigg[{  {\overline{y}} \over 5\,10^{-6} }\Bigg]\,
\Bigg[{ \theta_b\over 1.5'}\Bigg]^2 ~~~{\rm mJy}.
\end{equation}
Here the function $J(\xi)$ is related to $j(\xi)$ as described 
by Sunyaev \& Zel'dovich (1980);
angles  $\theta$ correspond to $w$ as given by $\theta = w/D_A$,
$D_A$ being the angular diameter distance;
the area averaged Compton parameter is given by
${\overline{y}}=2\,\pi\int_0^{\infty} \, d\theta
\,\theta\psi(\theta/\theta_b)
\,y(\theta)/\pi\theta_b^2$. For an individual structure the interesting
condition occurs when it
fills the aperture, that is, when
$\theta_b \mincir R/D_A$.

A related observable is provided by the  cosmic average of the Compton
parameter, that is, the integrated distortion of the CMB spectrum produced 
by all groups and clusters distributed in 
a given beam along the line of sight out to the
redshift $z$. This is given  by
\begin{equation} 
\langle y\rangle=
\int_0^z dz'\,{dV(z')\over dz'}\,\int dM\,\,N(M,z')\, \Bigg({R\over
D_A}\Bigg)^2
\int_0^{\infty} d\theta \, \theta \, \psi(\theta/\theta_b)\, y(M,\theta) ~,
\end{equation}
where $V(z)$ is the cosmological volume, and we take the mass
function $N(M,z)$ to have the Press \& Schechter form
that lies at the base of the all SAM descriptions of DM merging process.

Finally, the source counts corresponding to
fluxes larger than $S_{\nu}$ (Korol\"ev, Sunyaev \& Yakubtsev 1986) are
given by
\begin{equation}
N(>S_{\nu})=
\int_0^{\infty} dz\, {dV(z)\over
dz}\int_{\overline{M}(S_{\nu},z)}^{\infty} \,dM \,N(M,z)~,
\end{equation}
where $\overline{M}(S_{\nu},z)$ is the mass of a cluster (or group)
that at the redshift $z$ produces the flux $S_{\nu}$, computed after
eqs. (8) and (9).

\section{Results}

We plot in figs. 3 and 4 the relation $y-T$ that relates the Compton
parameter $y$ for individual groups or clusters to their ICM 
temperature.
The values provided by the self-similar scaling are   
shown by dotted lines, while dashed and solid lines 
represent what we predict for  moderate (case $\cal{B}$) and  
strong (case $\cal{A}$) feedback, respectively. 

Fig. 3 refers to the
critical
universe with $h=0.5, \,  \Omega_b= 0.06$ and with DM 
perturbation amplitude
$\sigma_8=0.67$ (SCDM cosmology/cosmogony hereafter). Fig. 4 refers to 
a flat universe with $\Omega_{\Lambda} = 0.7,
\Omega_{0} = 0.3, \, h = 0.7, \,  \Omega_b=0.04$ 
and $\sigma_8=1$ (in the following,
$\Lambda$CDM cosmology/cosmogony). In both figures the bottom panel shows
the $y-T$ relation with  $y$ normalized to the
self-similar
scaling $y_{grav}\propto T^{3/2}\rho^{1/2}$. The latter would
yield a horizontal line in this plot, so that departures from this
behaviour mark the effects of non-gravitational processes; it is seen that the
stronger is the feedback at group scales, the larger is the departure .

As to observability, note from figs. 3, 4 and from eqs. (9), (10) that the
difference $\Delta y=y-y_{grav}$ expected at the scale of
galaxy groups ($T \sim 1$ keV) is of order $5\, 10^{-6}$ or smaller,
corresponding
to  $\Delta T/T \mincir 25\, \mu$K. Observing such  values constitutes
a challenging proposition, but one becoming feasible for groups and poor
clusters
with size $R\sim 0.5\div 0.7\, h^{-1}$ Mpc at $z\sim 5\, 10^{-2}$ (and 
containing only weak radiosources), in the following instrumental 
conditions: last generation radiotelescopes (see Komatsu et al. 2000), with 
low internal noise and beams of a  few arcminutes, using long integrations
(the subtraction of atmospheric noise is favoured by the narrow
throws required for these objects); in the near  
future, arrays like AMiBA (see Lo, Chiueh, \& Martin 2000); 
in perspective, many superposed orbits of the planned {\it PLANCK}
 Surveyor mission within its apertures of $5'\div 10'$ 
(see De Zotti et al. 2000), and ALMA with its $\mu$ K sensitivity 
over angular scales from a few to tens of arcseconds (see www.ALMA.nrao.edu). 

The detection of $y$ in groups may  
be disturbed by surrounding large scale structures aligned along 
the line of sight. However, the probability for this to occur is low,   
since such structures compared with virialized 
condensations have lower temperatures and densities lower  
by factors  $\sim 30$ at least; only when 
their extension along the line of sight exceeds 
the size of a galaxy group by factors
larger than $30$ a comparable SZ signal will be produced.  
We add that the scatter we find in the $y-T$ relation 
is smaller than  $\Delta y/y=0.15$ for all temperatures; so
even in our moderate feedback case $\cal{B}$ 
the predicted {\it trend} of $y-T$ departs neatly  from 
$y_{grav}$ below a few keVs, and even upper bounds to $y$ may be of value.

In fig. 5 we plot (for both the SCDM and the $\Lambda$CDM
cosmology/cosmogony)
the predicted SZ source counts, and in fig. 6 we show also the contribution to 
the cosmic Compton parameter from all virialized structures distributed 
along the line of sight. Note 
that the difference bewteen the two feedback cases  
$\cal{A}$ and  $\cal{B}$ (solid lines and 
dashed lines, respectively) is small for such integrated observables 
(the curves corresponding to the 
self-similar case would closely overlap those relative to the 
moderate feedback case $\cal{B}$). 
This is because small and/or distant 
structures not filling a telescope beam contribute to these two 
observables with weights given by the system radius squared (see eq. 10-11); 
this circumstance underplays the
contribution of the groups, the structures 
most sensitive to feedback effects. 
On the other hand (see Cavaliere, Menci \& Setti 1991; Bartlett \& Silk 1994),
this very feature makes $N(>S_{\nu})$ suitable as a 
cosmogonical probe, since the contributions of the 
cosmogonical parameters like the amplitude $\sigma_8$ of the
spectrum of the initial DM perturbations override those 
from the state and evolution of the ICM. 

Similar results concerning $\langle y\rangle$ from the virialized 
structures have been found by Valageas \& Silk (1999)
(see also Valageas \& Schaeffer 2000). These authors add 
the contributions produced by the gas inside 
 lower density  structures and by truly intergalactic gas 
, 
which they find 
to depend sensitively on the heating model;  this is 
not unexpected from non-virialized gas that fills bigger volumes 
and can absorb larger energies. 

\section{Discussion}

Here we focus the basic features shared by a number of 
current models of the ICM (MC2000, 
CGM2000; Wu, Fabian \& Nulsen 1998; Valageas \& Silk 1999; 
Bower et al. 2000),  
in order to discuss to what extent our predictions concerning 
the SZ effect are robust. 
We first stress the basic blocks 
shared by such models. Next we discuss  
our specific treatement of the boundary conditions 
in terms of shock jumps. 
Then we discuss why the shocks are effective in amplifying the impact of 
non-gravitational energy injections. Finally, we show how 
the departure of the  $y-T$ relation from its gravitational 
counterpart is related to the corresponding departure of the $L-T$ relation;  
the model-independent link is constituted by 
the entropy discharged by the injection of 
non-gravitational energy, but the actual size of such departures depends 
on the feedback strength. 

\subsection{Basic blocks}

a) Part of the baryons are subtracted out of the hot phase by cooling and
are locked in the stars formed, to be only in part recycled by stellar
winds.
 
This actually constitutes
a minor negative contribution 
to the mass of the hot phase. 

b) The energy fed back by the condensing baryons, either to fuel black holes 
at the galactic centers or to form massive stars followed by winds and SNe, 
heats up a fraction of the intergalactic medium. In our model 
the energy injection is $E_* \approx 3\,
10^{48}$ erg per solar mass of stars formed, the fraction is 
$f_* \mincir 0.15$ and the stellar heating temperatures are 
$kT_* \sim 0.2$ keV; 
the net outcome is to produce in groups
a central density $n\sim 10^{-4}$ cm$^{-3}$, lower than in clusters.
Since such gas is
recovered by larger haloes during later merging events (with the delays
described in \S 3.1), the effects propagate a number of steps up the mass
hierarchy, so that the $L-T$ relation departs away from $L_{grav} \propto
T_V^2$ at temperature considerably higher than  $0.5$ keV. A 
corresponding feature is 
also present, e.g., in the elaborate model used by Bower et al. (2000); 
in their model, as well as in ours, the gas fraction inside the virial 
radius is lower in groups than in clusters.

c) Equilibrium is re-established soon after minor and intermediate 
merging events, that is, over sound crossing times which are somewhat 
shorter than the structures' 
dynamical times. In such conditions, 
the pressure gradient balances gravity and a 
$\beta$-model holds.

d) Once equilibrium is assumed, a key role is played by the boundary 
conditions that set normalization and shape (specifically, the values
of $\beta$) of the ICM density profiles. In particular, the pressure 
balance at the boundary $r \approx R$ clearly constitutes the key condition; 
we add that such a balance is to include the 
ram pressure due to the bulk motion $v_1$ of the infalling 
gas out of equilibrium, to read $p_2 \approx  p_1 + n_1 m_H v_1^2$ 
in conditions with low bulk kinetic energy downstream.

e) Since  the ICM state is observed to be close to isothermal ($T$ may 
decrease out to $R$ hardly by a factor 2, see Nevalainen, 
Markevitch, \& Forman 2000) a sharp boundary -- that is, some sharp drop or 
discontinuity of $T$ and/or $n$ -- is to take place at $r \approx R$, 
lest the thermal
exceeds the integrated gravitational energy.

Model-dependent features, which however 
affect to a lesser extent the profiles 
of the ICM in virialized structure, include: 
the exact value of the polytropic index (close to 1 anyway) and 
the detailed form of the underlying DM potential  (see the discussion 
by Cavaliere, Menci \& Tozzi 1999); the parameters other than 
the feedback exponent $\alpha_h$ that are introduced in the various SAMs 
only to tune the optical luminosity
function and the Tully-Fisher relation for the galaxies 
(cf. Cole et al. 1994 with Cole et al. 2000).

\subsection{Large-scale shocks}

It is also widely agreed that shocks provide an effective coupling of the
gravitational to the thermal energy in the ICM.

In our picture, we take up from the results of the simulations 
 referred to in \S 3.2 
the notion that {\it external}
accretion
shocks do establish a discontinuity at about $r \approx R$, with low bulk
velocities left downstream; observational evidence of large scale shocks
related to violent merging
events is reported by Roettiger, Stone \& Mushotzky (1998); Roettiger, Burns
\& Stone (1999); Markevitch, Sarazin \& Vikhlinin (1999). Across such shocks
the above stress conservation combined with with mass conservation yields the
classic Rankine-Hugoniot jump conditions for $T$ and $n$.
Without further ado these produce a non-linear increase of 
$G \equiv n_2/n_1$ from  $1$ to $4$ as the infall flows become increasingly 
supersonic, that is, as $T_V/T_1$ increases. 

Note that the merging events that
yield the largest contributions to $G(T_2/T_1)$ are the numerous ones 
(more than 90\%) with minor partners  having mass ratios smaller than $1/4$, 
which may be appropriately described as constituting an 
``accretion'' inflow of the 
DM component as well as of the baryonic component (see Raig, Gonzalez-Casado 
\& Salvador-Sole 1998). 
Such minor events contribute
the majority of the mass increase (more than 50\%), and imply lower 
temperatures 
$kT_1 \mincir 1 $ keV (whether of gravitational or of stellar origin) 
for the external gas they involve, leading to more supersonic 
gas inflows. The summed action of these 
events is as close to isotropic as permitted by the surrounding 
large scale structure (see Tormen 1997; Colberg et al. 1999); 
re-establishing the equilibrium 
then is more like a continuous re-adjustement to the accretion inflow.

On the other hand, given the external conditions
 the gas inflow is bound to be more 
supersonic on average when it  takes place 
into the deeper wells provided by the richer clusters where $v_1 \magcir
1000$ km s$^{-1}$ is attained; 
this leads
to expect larger entropy depositions in the outer regions of rich clusters,
as in fact observed (Lloyd-Davies, Ponman \& Cannon 1999). In poor groups,
instead, where the stellar heating temperatures match the virial ones to give 
$kT_* \sim 0.3-0.5$ keV, the shocks degenerate into sonic, adiabatic 
transitions; here all internal densities scale down proportionally to 
$n_2\rightarrow n_1$, so $L
\propto n_2^2$ is bound to plunge down toward $L \propto n_1^2 \propto 
n^2_2/16$.

To see how the departures from the self-similar scalings begin and end,
one can make
use of two opposite approximations, rather crude but simple: when the
pre-shock
temperature $T_1$ is low,  the strong shock approximation yields $kT_2
\approx kT_V +
3\, kT_1/\, 2$, see eq. (6); on the other hand, the
lower bound to all temperatures is constituted by $T_*$. For decreasing
depths
of the wells these approximations show that the decline of $\beta \propto
T_V
/(T_V + 3T_1/2)$  begins slowly, then becomes faster like $\beta \propto
T_V/T_*$ in shallow wells; this implies correspondingly flatter density
profiles. Similarly but more strongly, $L \propto n^2_1\, g^2(T/T_1) \approx
16
n^2_1\, (1- 15\, T_1/8\, T_2)$ departs from $L_{grav} \propto 16\, n_1^2$
slowly at first in deep wells, but eventually plunges toward  
$L \propto n^2_1$ in shallow wells.

A final remark concerns the normalization of the 
Compton parameter $y$ related to the shock condition.
In our model, the boundary values for the density and the temperature 
are fixed by the shock jumps in terms of the external density $n_1$ 
and of the 
gravitational potential $\phi_2$ at $r= R$ (see eq. 7).
Now, $n_1\propto \Omega_b$ holds,  
while $\phi_2$ depends both on the shape of the gravitational potential 
and on the location of the boundary at $r\approx R$. 
For our fiducial choice of $\Omega_b=0.04$ and of the Navarro et al. (1995) 
potential, our  shock or boundary location  
(supported by the N-body simulations referred to in \S 3.2) leads 
to a normalization of the 
$y$ parameter in agreement with the observed values; for example, 
for a Coma-like cluster with $T_V\approx 8$ keV, we obtain 
$y\approx 10^{-4}$, closely independent of the feedback, see fig. 4; 
this is in good agreement 
with the measured value, see Herbig et al. (1995). 
The same normalization can be obtained on changing the value of $\Omega_b$ in 
the range $0.02-0.06$ allowed by the canonical primordial 
nucleosynthesis, and on retaining the same 
DM potential but readjusting the shock 
position within the range $0.85-1.15\;R$    
compatible with the N-body simulations. 

\subsection{Amplification and feedback} 

It is intrinsic to the present model that the  departures from
the gravitational scaling  are non-linearly {\it amplified}, due to the varying
shock strengths set by the depth $k T_V$ of the potential wells compared
with $kT_*$. For a given departure, such an amplification {\it lowers} the
requirements for the non-gravitational energy injection.
Indeed, the excess central entropy $e^S \propto k T/ n^{2/3}$ 
observed in groups over the self-similar expectations (see fig. 2) 
can be obtained with a moderate energy injection $E_*$ leading to 
$T \approx T_V+T_*$ with $k\, T_* \approx  0.3$ keV, provided the 
central density at the group scale is decreased relative to clusters 
by a factor around $10$. 
In our model such a decrement results in part from the boundary density jump 
(lower than in clusters by a factor 4), and in part 
from the shallow gas density profile due to the lower values 
of $\beta$ (about half than in clusters). 

Thus, the presence of shock amplification allows us to consider 
{\it conservative} quantities for the feedback, namely, 
energy contributed by stellar winds plus by SN explosions adding up to 
$E_* = \eta_{SN}\,
5\,10^{50} \approx 3\, 10^{48}$ erg per $M_{\odot}$ of stars formed. 
Note also that the
two contributions are largely separated and undergo different amounts of
cooling, lesser (if anything) for the winds.
 
Stellar winds and amplification go a couple of steps toward meeting the
concerns about prompt radiative cooling of the SN outputs which may limit 
the fraction available for action at galactic or group scales. 
As is well known, the issue is
a thorny and debated one. On the one hand stands the intrinsic sensitivity of
the cooling to density clumpiness (see, e.g., Thornton et al. 1998). On the
other hand, momentum and thermal energy are carried outward of the sites of 
star formation in the form of
blast waves and shocks, in fact as SN Remnants. It is widely agreed
(see Ostriker \& McKee 1988) that these may easily overlap before they 
enter a strong radiative phase, especially when they are driven by type 
II SNe produced by correlated massive star
formation, and when they are helped by the tunnel network in the ISM. 
So they merge  their hot and thin interiors, break out of the galaxies 
in the form of superbubbles, and 
discharge their high entropy content into the galaxian outskirts; 
this is then mixed throughout the ICM. 

Additional sources of non-gravitational heating may be provided by 
Active Galactic Nuclei shining at the center of developing galaxies
(see Wu, Fabian \& Nulsen 1999; Valageas \& Silk 1999; 
Aghanim, Balland \& Silk 2000). 
The AGNs are fed at sub-pc scales by accreting
black holes; 
their outputs are huge, but the effective coupling within galaxies or groups 
of such energy to the ICM within galaxies or groups 
is even more uncertain than for the SNe. Such coupling 
would require quite some tuning to yield just the observed $L-T$ correlation 
(P. Madau, private communication); the intrinsic 
instability of the cooling may 
easily drown any correlation into a large intrinsic scatter.
Treating such processes in full is beyond the 
scope  of the present paper and will be undertaken elsewhere. 

\subsection{SZ effect vs. X-ray emission} 

As can be noted from figs. 3 and 4, the departure from self-similarity 
we find from our model in the $y-T$ relation for groups and clusters  
features an 
{\it inverse} $y$-dependence on the strength of the feedback, which sets in 
below $T\sim 1$ keV. 
This behaviour is similar to that of the $L-T$ relation 
for groups and clusters of galaxies. 

The correspondence is explained by the model-independent relation which
follows from combining the definitions  of $S$ and 
of the Compton parameter $y$
with its self-similar scaling given by  by eq. (3), to read
\begin{equation}
e^{S- S_{grav}} = (y/y_{grav})^{-2/3}\,(T/T_V)^{5/3} ~.
\end{equation}
We recall from \S 5 that $T$ exceeds $T_V$ due to the
non-gravitational heating, and the excess becomes more and more relevant 
in moving from rich clusters to poor groups. 
The above eq. shows than that an internal heating source rising the 
entropy content will result in lower values of $y$ 
compared to the self-similar expectations at the scale of groups. 

The counterpart of eq. (12) for the X-ray emission reads 
\begin{equation}
e^{S- S_{grav}} = (L/L_{grav})^{-1/3}\,(T/T_V)^{7/6}~, 
\end{equation}
again model-independently. Assuming the two observables 
are associated (albeit with different shape factors) 
with ICM confined within comparable sizes, 
the two equations combine to yield 
\begin{equation}
y/y_{grav} = (L/L_{grav})^{1/2}\,(T/T_V)^{3/4}~. 
\end{equation}

The above equations highlight in a {\it model-independent} way 
the inverse nature of both the $S-y$ and 
the $S - L$ relations.  
So the entropy excess observed in groups of galaxies implies a
related deficit of {\it both} the  X-ray luminosity as indeed is found 
in the observed $L-T$ correlation, and of the SZ effect as here predicted. 
In both cases the dependence 
on the temperature is instead direct, and actually stronger 
in the $S-y$ relation.

The point to emphasize is that 
these two probes are  {\it independent}  
observationally; in fact, they are
measured in such distant bands as X-rays and microwaves, with very different
instrumentations subject to different 
systematics.  In addition, the
SZ effect can be measured also at sub-mm wavelengths, which provide 
a positive $\Delta I/I$ with systematics different yet. 
Finally, the selections are intrisically different in the $\mu$w/sub-mm and 
in the X-ray bands, with the SZ signal being much
{\it less} sensitive than the X-ray emission to internal density 
clumpiness or enhancements. 
Thus, while either observable may be subject to its own 
observational biases, background subtraction or scatter 
(see Mahdavi et al. 1997; Roussel, Sadat \& Blanchard 2000), 
the {\it combined} evidence
will be highly significant and strongly constraining 
for the non-gravitational heating mechanisms. 
 
\section{Conclusions}

In this paper we have computed and presented three observables related to
the SZ effect from the ICM in groups and clusters of galaxies, namely, 
the $y-T$ correlation, the source counts and the contribution to the 
cosmic SZ effect. We
based our computations on the specific semi-analytic model (SAM) described in \S 3;   
this is built upon the 
hierachical merging histories of the DM component of such structures, and 
provides stellar and X-ray observables in agreement with the data.
The picture underlying our
SAM is widely shared, and its main blocks are common 
(as discussed in \S 5) 
to other SAMs also aimed at explaining the optical and X-ray observations. 

One key feature is the energy and momentum fed back into the ICM by the
condensing baryons. 
This has the effect of pre-heating the baryons that 
fall into groups 
and clusters during their merging history. The shock front forming 
between the 
infalling and the internal gas amplifies the effect of the 
non-gravitational heating on the 
density distribution of the gas in a way which depends on the cluster or group 
temperature $T$. The net effect is to 
break the (approximate) scale-invariance 
of the DM quantities in {\it virialized} structures, particulary in 
galaxy groups with $kT\mincir 1$ keV compared to rich clusters. 
The effect of such processes on the X-ray properties of
groups and clusters has been investigated
 in our previous papers; here we have shown that 
the SZ effect constitutes a {\it complementary} 
probe of the non-gravitational 
heating of ICM. In particular we emphasize the following results. 

$\bullet$ The measure of the Compton parameter $y$ from deep, targeted observations 
of groups and clusters and the resulting $y-T$ correlation provide  
a direct probe of the energy discharged into
the ICM and archived there in conditions of long cooling times.
Specifically, for energy injections dumped into the ICM 
within galaxies or groups we predict the $y-T$ correlation 
to bend down systematically relative to the self-similar expectations
(see \S 4, figs. 3-4). 
In \S 5 we discuss the generic aspects 
of the {\it trend}: larger injection - smaller $y$ 
at group scales, that  we expect from any reasonably complete model.  
The level of such departures will be 
similar to ours for feedback models able to 
reproduce the bent shape of the $L-T$ relation in X-rays.
This conclusion is substantiated by 
eqs. (12)-(14) which show {\it model-independently} that heating sources 
rising the internal entropy in groups to the observed ``floor'' 
will result in observably low values of $y$ compared to the self-similar 
expectations. 

$\bullet$ The integrated SZ effects from structures (i.e., 
the counts $N(> S_\nu)$ of SZ sources associated with 
virialized structures and their contribution 
$\langle y\rangle$ to cosmic SZ effect (see figs. 5-6) 
are dominated by clusters less affected by the feedback, even less  
by its details.  

To conclude, the emerging picture is one where the close 
scale-invariance of the {\it gravitational} energy released over large scales 
under the drive of the DM dynamics is broken by the {\it nuclear} 
energy that at small scales drives the stellar life and death; the 
break-even point should occur at the transitional mass scale $M\sim 
10^{13} M_{\odot}$ where $kT_V \approx kT_* \approx 0.3$ keV 
holds. The specific way in which baryons break the DM 
scale-invariance can be probed with several {\it complementary} 
but 
observationally {\it independent} observables which include: the bent 
X-ray luminosity-temperature correlation; the weak or absent evolution 
of the X-ray luminosity functions; excess counts of the ICM X-ray 
sources correlated with high rates of cosmic star formation at 
$z\magcir 1.5$; the shape and evolution of the optical luminosity function 
of the galaxies; finally, the SZ effect as proposed here, and in 
particular
the bent $y-T$ relation from clusters to groups.

\smallskip
\noindent
{\bf Acknowledgements}. This work benefited at all stages from helpful
exchanges with G. De Zotti. We are grateful to F. Melchiorri for
discussions of his observational strategies with MITO, to A. Lapi 
for critical reading, and to 
our referee for constructive and helpful comments. Partial grants from 
ASI and MURST are acknowledged.

\newpage
\section*{FIGURE CAPTIONS}

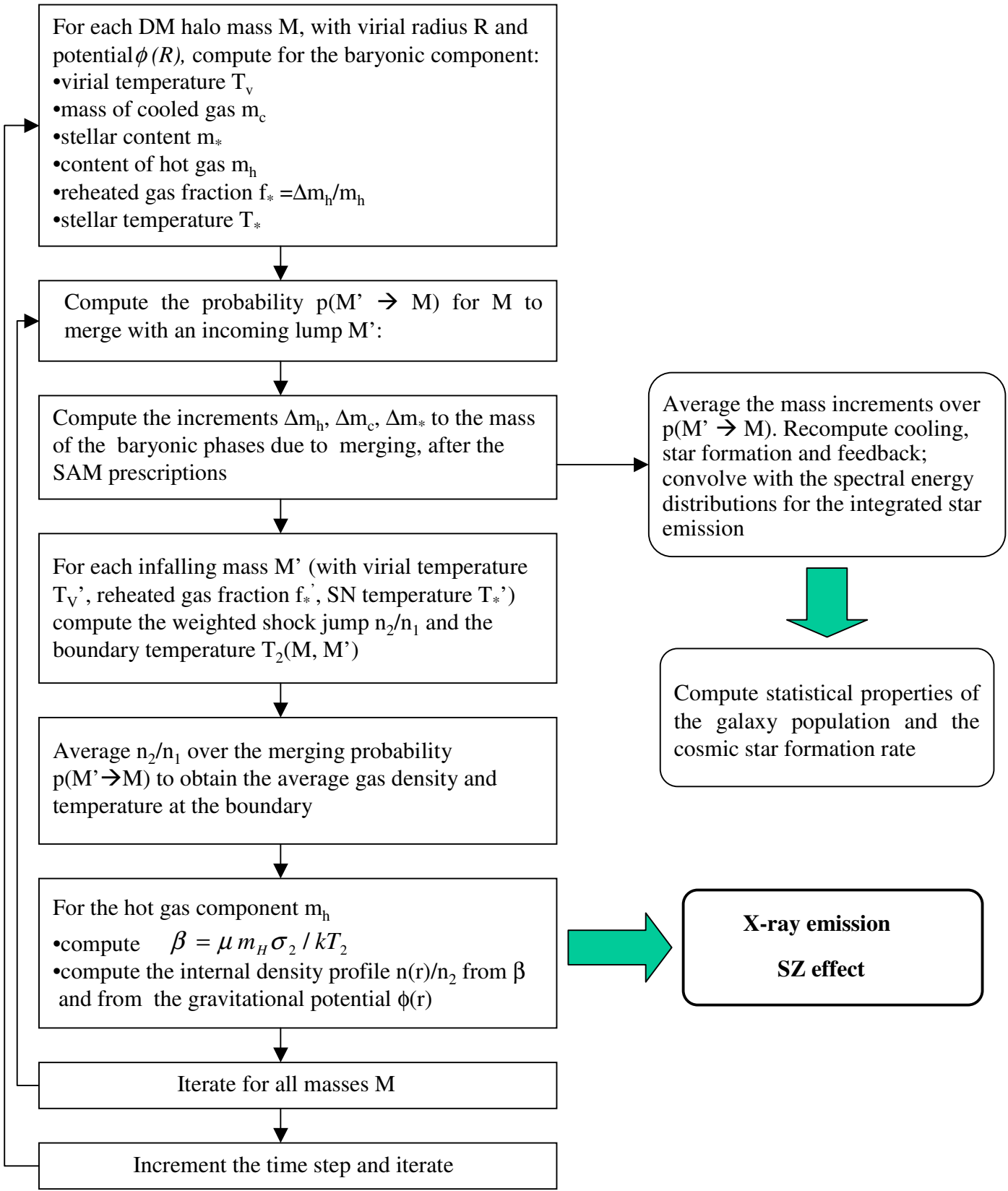
Fig. 1. We show in the form of a flow chart diagram the processes 
we consider and the computational steps we use to compute 
in the DM haloes the 
ICM profile and its boundary conditions. 

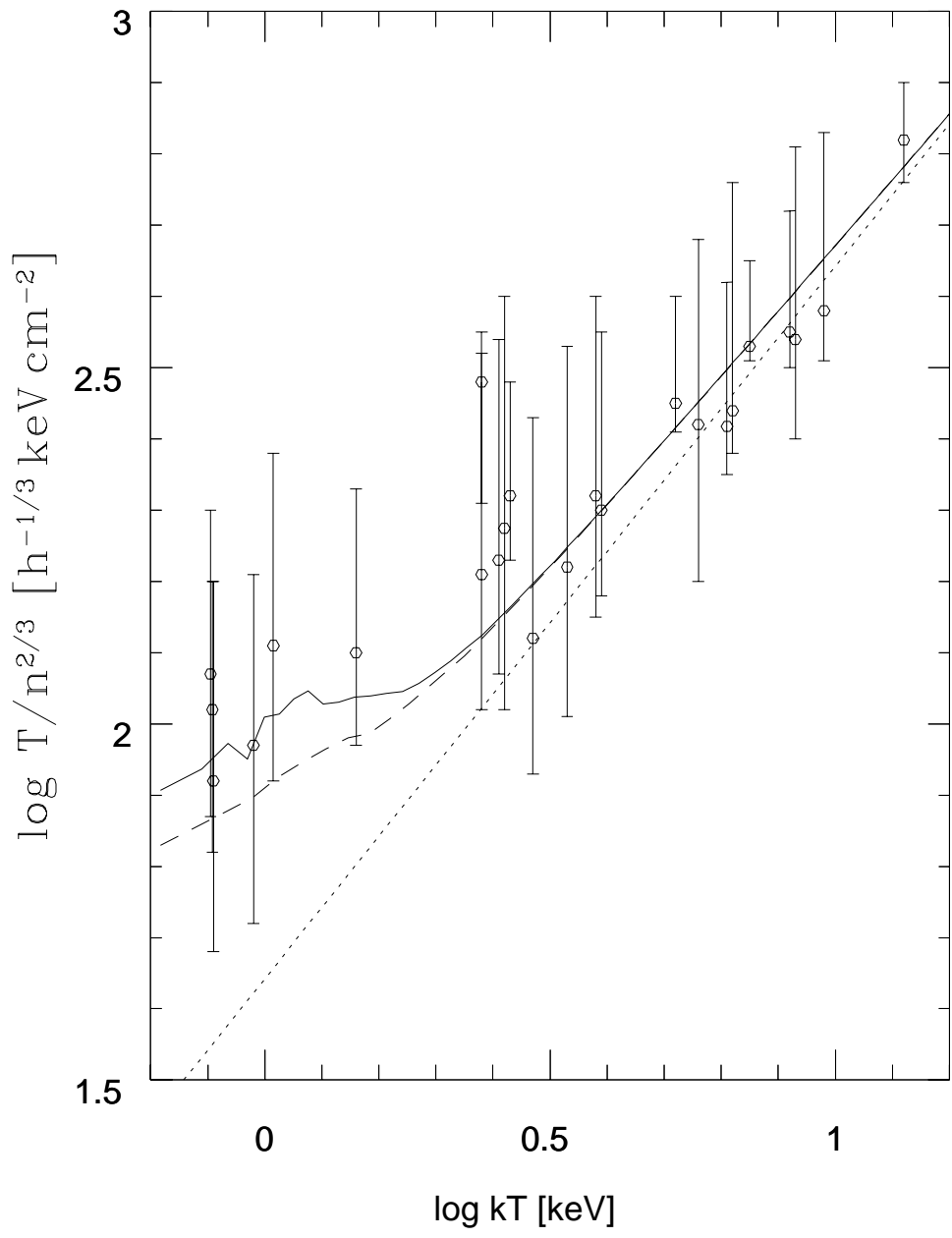
Fig. 2. As a preliminary test of our SAM model,
we show the predicted central entropy of the ICM 
as a function of the temperature for the 
strong feedback case $\cal{A}$ (solid line) and the moderate feedback case
$\cal{B}$ (dashed line), compared
with the data from Ponman, Cannon \& Navarro (1999). We also
show as a dotted line the result from the self-similar scaling, 
eq. (2). All lines refer to the SCDM cosmogony/cosmology with $\Omega= 1$,
$\Lambda=0$, $h = 0.5$,  $\Omega_b= 0.06$, and $\sigma_8 = 0.67$. The
$\Lambda$CDM case yields similar results.

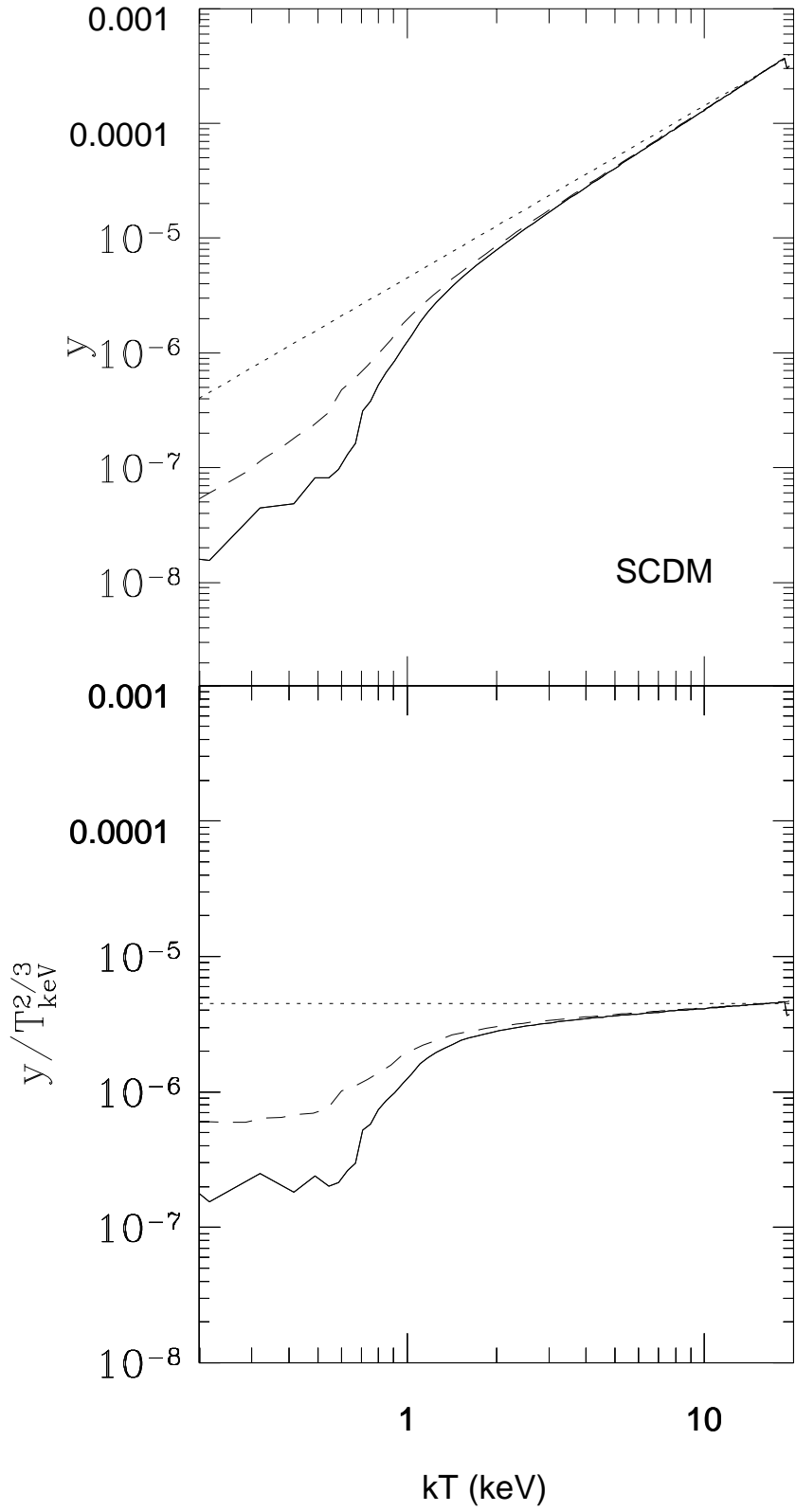
Fig. 3. The top panel shows the relation we predict  
between the (area averaged) Compton parameter $y$ for individual  
groups and clusters and the 
temperature; the solid line refers to the strong feedback
case $\cal{A}$, the dashed line to the moderate feedback case $\cal{B}$,
while the self-similar scaling is plotted as a dotted line. 
SCDM cosmology/cosmogony. The bottom panel shows
the same correlation but with the Compton parameter normalized to its
self-similar scaling given by eq. (3)

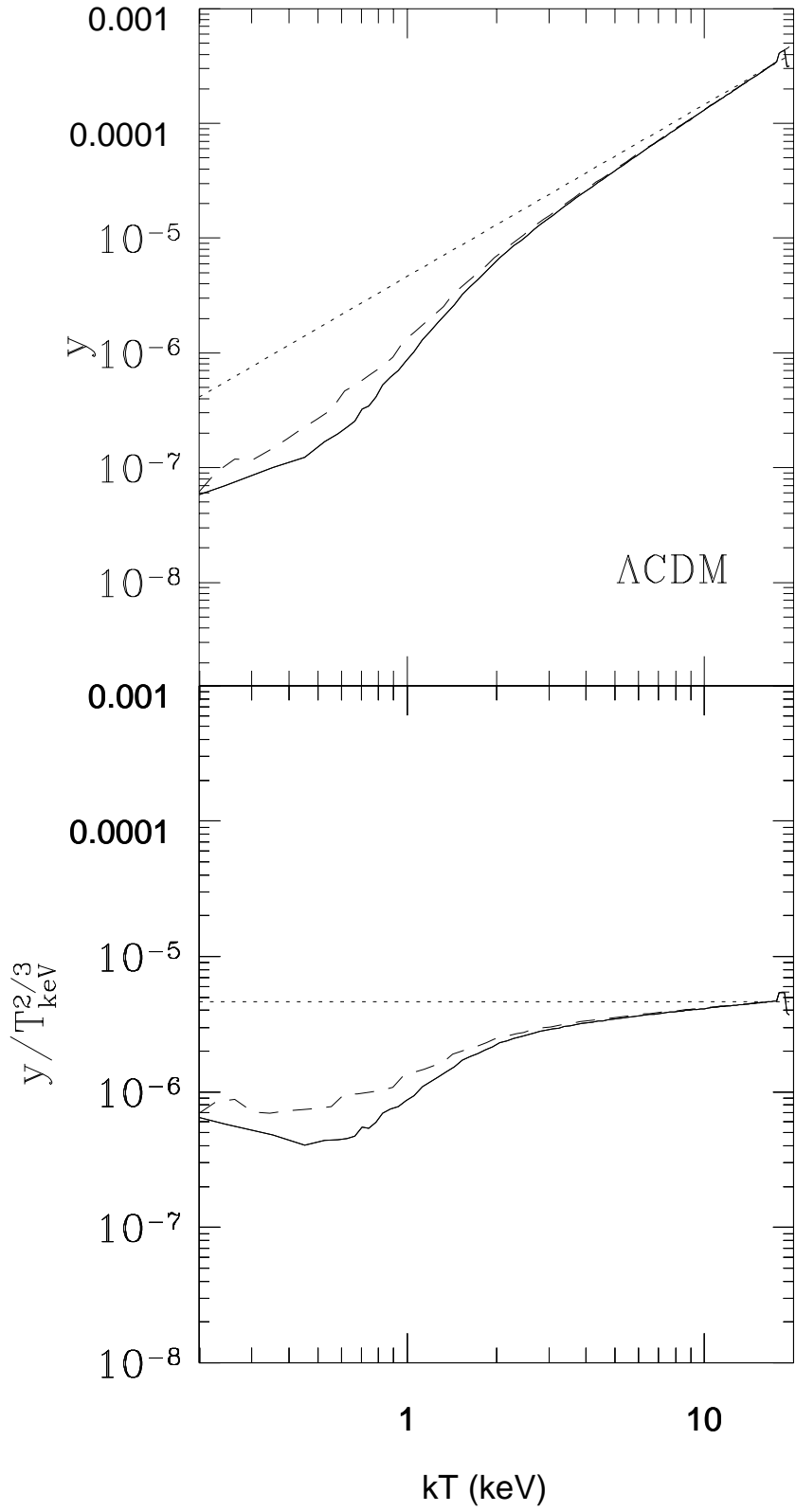
Fig. 4. Same as fig. 2 but for the $\Lambda$CDM cosmogony/cosmology.

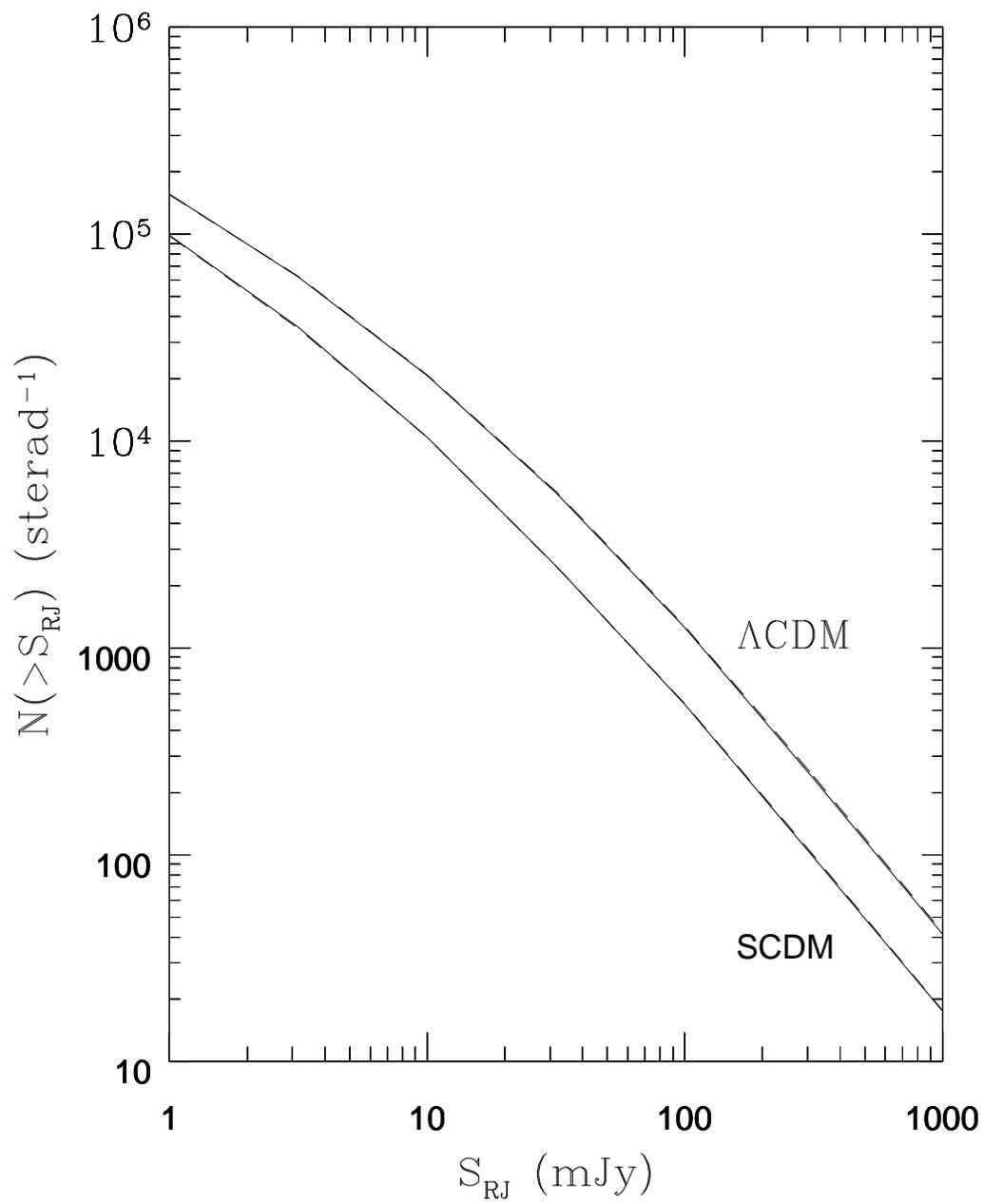
Fig. 5. The predicted source counts as a function of the SZ flux at
100 GHz, see eqs. (10) and (11). Solid and dashed lines refer to the
feedback cases $\cal{A}$ and $\cal{B}$, respectively. 
The curve  relative to the self-similar case would overlap that
referring to case $\cal{B}$. The bottom line is
computed for the SCDM cosmogony/cosmology, while the one on top refers to
the
$\Lambda$CDM case.

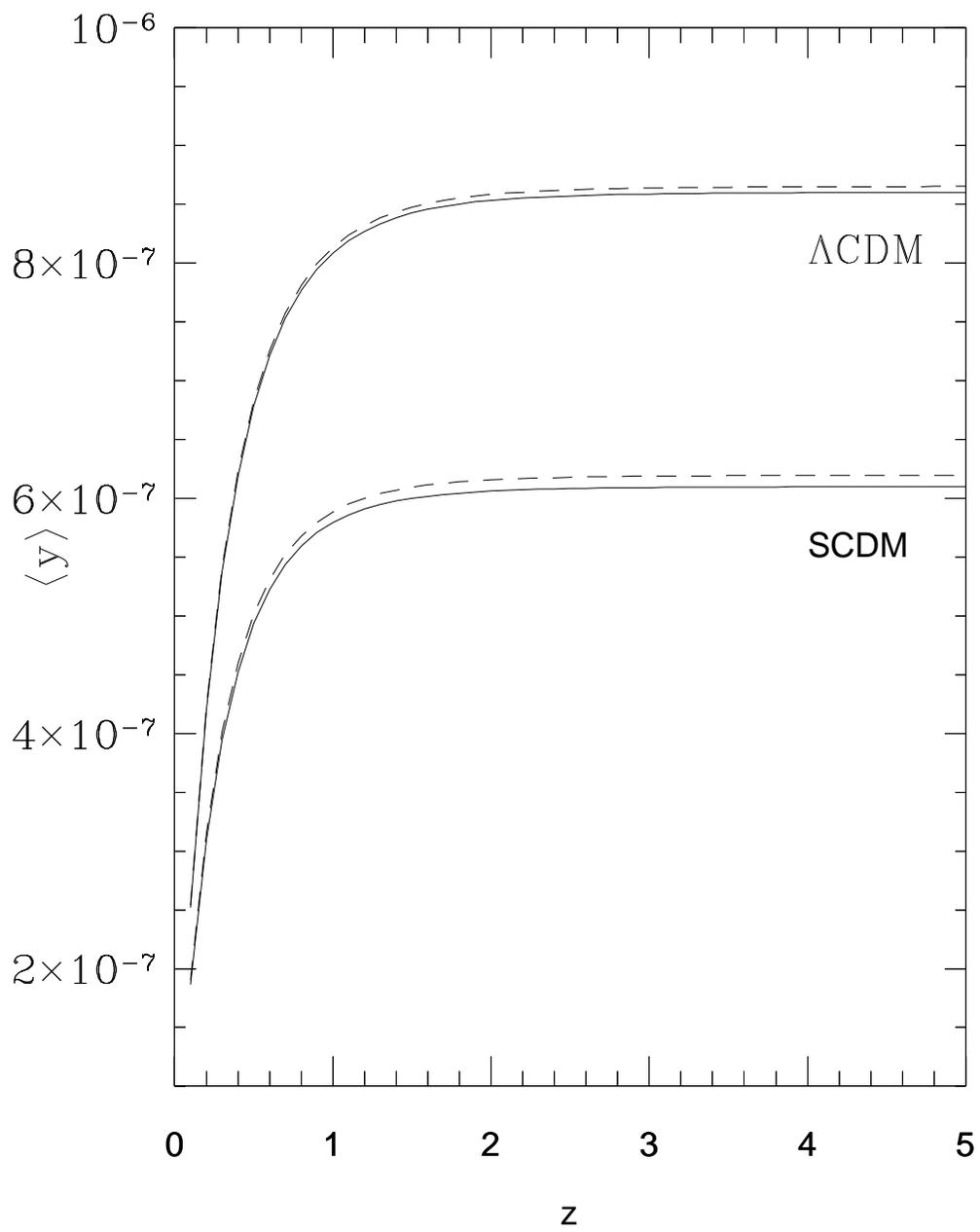
Fig. 6. The contribution to the cosmic Compton parameter (see eq. 11) 
from the ICM inside {\it virialized} structures, groups and clusters of 
galaxies, distributed out to the redshift z; SCDM (bottom lines),  
and $\Lambda$CDM universe (top lines) are considered. 
The solid and dashed lines refer 
to the feedback strengths $\cal{A}$ and $\cal{B}$, respectively.
\end{document}